\documentclass[12pt,english]{article}

\usepackage{graphicx,xcolor,subcaption,babel}
\usepackage{natbib}

\date{}

\usepackage{authblk}

\title{
{\begin{center} \fontsize{13}{13}\selectfont  
{\bf White Paper Submitted to ``Decadal Survey for Solar and Space Physics 
(Heliophysics) 2024-2033''} \end{center} }\vskip 40pt
{\fontsize{17}{17}\selectfont \bf Regular Solar Radio Imaging at Arecibo: 
Space Weather Perspective of Evolution of Active Regions} }

\author[1]{\fontsize{13}{12}\selectfont Periasamy~K.~Manoharan}
\author[1]{\fontsize{13}{12}\selectfont Christopher~J.~Salter}
\author[1]{\fontsize{13}{12}\selectfont Christiano~M.~Brum}
\author[2]{\fontsize{13}{12}\selectfont Stephen~M.~White}
\author[3]{\fontsize{13}{12}\selectfont Phil~Perillat}
\author[3]{\fontsize{13}{12}\selectfont Alfredo~Santoni}
\author[3]{\fontsize{13}{12}\selectfont Felix~Fernandez}
\author[4]{\fontsize{13}{12}\selectfont Tapasi~Ghosh}
\author[1]{\fontsize{13}{12}\selectfont Benetge~Perera}
\author[3]{\fontsize{13}{12}\selectfont Arun~Venkataraman}

\affil[1]{\fontsize{11}{10}\selectfont Arecibo Observatory, University of Central Florida, 
Arecibo, Puerto~Rico~00612, USA.}
\affil[2]{\fontsize{11}{10}\selectfont Space Vehicles Directorate, Air Force Research 
Laboratory, Albuquerque, NM, USA.}
\affil[3]{\fontsize{11}{10}\selectfont Arecibo Observatory, Yang Enterprises Inc.,
Arecibo, Puerto~Rico~00612, USA.}
\affil[4]{\fontsize{11}{10}\selectfont Independent Astronomer, Arecibo, Puerto~Rico~00612, USA.}

\begin{document}{}
\maketitle

\begin{abstract}
The sudden release of magnetic energy on the Sun drives powerful solar flares 
and coronal mass ejections. The key issue is the difficulty in predicting the 
occurrence time and location of strong solar eruptions, i.e., those leading to 
the high impact space weather disturbances at the near-Earth environment. Solar 
radio imaging helps identify the magnetic field characteristics of active regions 
susceptible to intense flares and energetic coronal mass ejections.  Mapping of 
the Sun at X-band (8.1 -- 9.3 GHz) with the 12-m radio telescope at the Arecibo 
Observatory allows monitoring of the evolution of the brightness temperature of 
active regions in association with the development of magnetic complexity, which 
can lead to strong eruptions. For a better forecasting strategy in the future, such 
ground-based radio observations of high-spatial and temporal resolution, along with 
a full polarization capability, would have tremendous potential not only to 
understand the magnetic activity of solar eruptions, but also for revealing the 
particle acceleration mechanism and additional exciting science.

\end{abstract}

\newpage
\pagenumbering{arabic}
\setcounter{page}{1}

\section{\fontsize{14}{14}\selectfont Introduction}

The magnetic field in the solar atmosphere essentially controls the plasma structure, 
storage of free magnetic energy, and its release as flares and/or mass ejections. The 
areas of strong magnetic field concentration on the surface of the Sun form active 
regions, which are embedded in a group of sunspots of the same magnetic polarity,
followed by a group of sunspots of opposite polarity. They display closed magnetic 
field lines resulting from the bipolar geometry, as well as complex field structures 
given by the mixing of fields. An extensive record of sunspot activity shows that 
the location of sunspot formation on the solar disk and their number count change 
with time. The direction of the magnetic field associated with sunspots reverses 
polarity between hemispheres in a period of $\sim$11 years. In the 
subsequent 11-year period, the orientation of the magnetic fields in the northern and 
southern hemispheres of the Sun restores, taking about 22 years to complete a full 
solar magnetic cycle (e.g., \citealt{hathaway2015}). In each 11-year period, 
as the magnetic fields change, the amount of magnetic activity above the surface of 
the Sun, estimated by the number of sunspots, also increases to a peak and then 
decreases. At the same time, the number and size of sunspots, solar flares, coronal 
mass ejections (CMEs), the levels of solar radiation and coronal structures exhibit 
synchronized variations with the phase of the 11-year solar cycle. The rate of 
occurrence of flares and CMEs maximizes towards the peak of the solar cycle 
(e.g., \citealt{lamy2019}). From the viewpoint of space weather, energetic CMEs 
create disturbances in the entire heliosphere, driving shocks and accelerating 
electrons and protons, evidenced by radio bursts and solar energetic particle 
events.

\section{\fontsize{14}{14}\selectfont Solar Active Regions and Space Weather Conditions}

Flares and CMEs are energetic phenomena, involving bursts of electromagnetic radiation 
and dynamical eruption of plasma and magnetic field from active regions, caused by the 
magnetic reconnection process (\citealt{priest2002}). In the context of space weather, 
intense flares and their associated CMEs, particularly those directed towards the Earth, 
are important, because they drive the large-scale energetic space-weather storms in the 
interplanetary space and give rise to hazardous effects at the near-Earth environment 
(e.g., \citealt{xie2006}; \citealt{pankaj2011}; \citealt{mano2018}). The essential need 
of space-weather forecasting is to 
monitor the evolution of an active region, as it rotates close to the central meridian 
of the Sun, and assess its likelihood of releasing an Earth-directed solar flare and/or 
CME (e.g., \citealt{mano2005}). The X-ray and EUV emissions from the optically-thin 
corona above an active region, originating at the top of the complex magnetic field 
network, relate to the inhomogeneous, hot, and over-dense plasma and they provide 
a diagnostic of the magnetic activity above the active region (\citealt{sam2008}). 

Specifically, active regions coupled to the sunspot groups of complex polarity 
$\delta$-spot configuration, as per the Hale or Mount Wilson scheme 
(e.g., \citealt{hale1919}; \citealt{kunzel1965}), are prone to produce 
significantly intense flares. However, such spots are limited in number to only 
$\sim$4\% of the total number of spots compared to the numerous $\beta$ spots of 
bipolar characteristic, which produce flares of much lower intensity. Along with 
the number of sunspots, as the solar cycle 
progresses, the latitude of sunspots move gradually towards the solar equator.
However, the formation of a complex or a simple active region shows no significant 
preference of latitude and originates from the same reservoir of flux in the solar 
interior (\citealt{jaeggli2016}). Thus the continuous monitoring of development 
of complexity of an active region is indispensable to achieve a reliable predictive 
capability of solar eruptions, and their effects at the near-Earth space in the context 
of space weather. 

\begin{figure}[!t]
\begin{center}
\includegraphics[height=7.7cm]{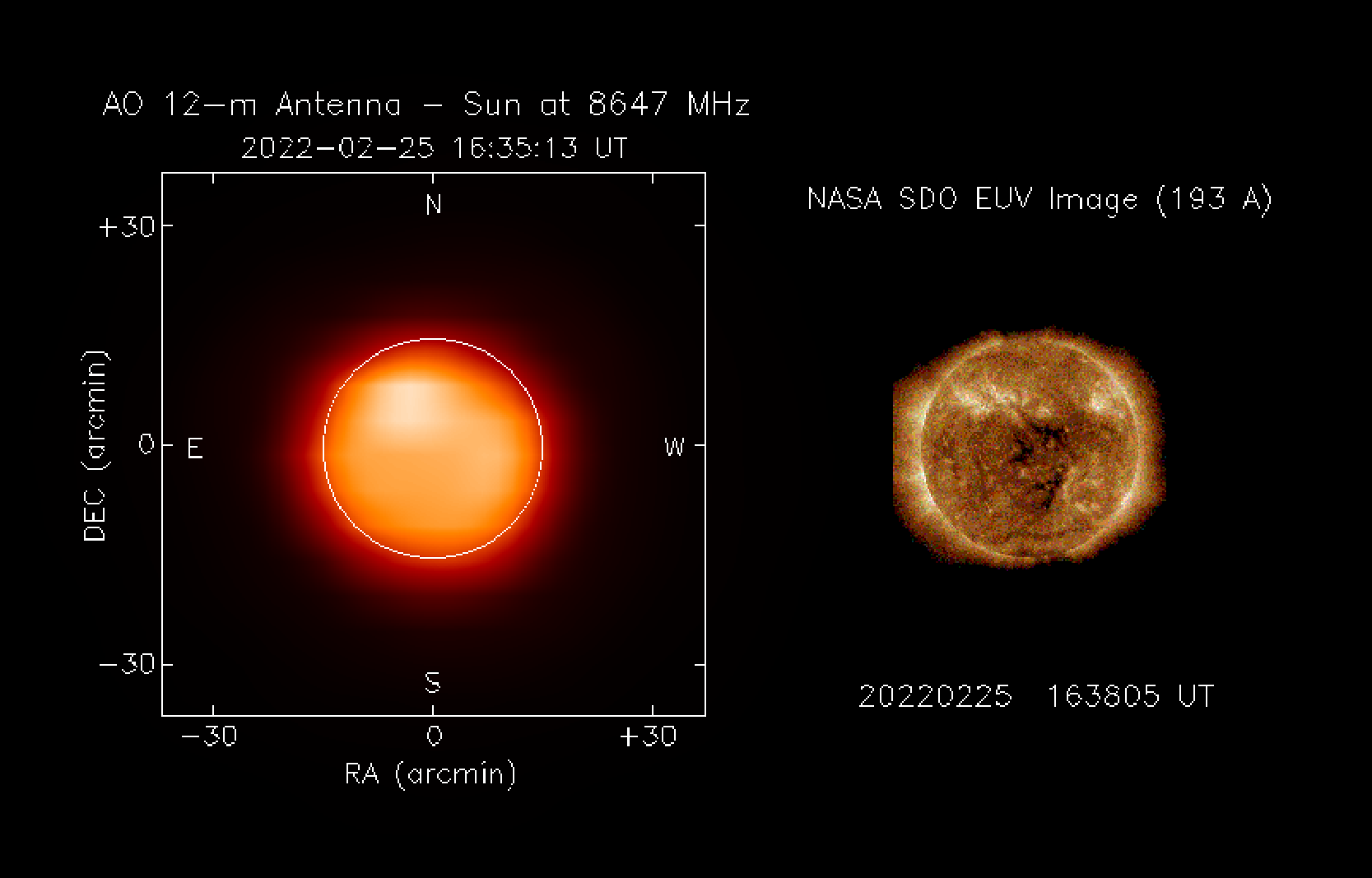}
\end{center}
\textbf{Figure 1:} A sample image of the Sun on 25 February 2022 at 8647 MHz made with 
the 12-m radio telescope at the Arecibo Observatory (left). The intense emitting 
active regions and the coronal hole of quiet emission are clearly seen in the image. 
For comparison, a near-simultaneous image of the Sun observed with the AIA/SDO in the 
193~\AA~wavelength band is shown at the right. 
\end{figure}

The X-ray and EUV brightness resulting from the magnetic reconnection process serves as 
a scale to gauge the energy release mechanism in the corona 
(e.g. \citealt{kosugi1997}; \citealt{priest2002}; \citealt{sam2008}). 
In a similar way, the radio signatures from an active region in the frequency range of 
5 -- 10 GHz are the gyro-synchrotron radiation from high-energy electrons trapped in 
small-scale magnetic field loops and the observed bright features are gyro-resonance 
emitting regions where the field strength exceeds 600 G (e.g., \citealt{bastian1998}; 
{\citealt{white1999}; \citealt{nindos2020}). Typically, the gyro-resonance spectrum peaks 
in the range 5 -- 10 GHz and corresponds to the transition region and provides a direct 
measure of magnetic fields above the photosphere (\citealt{gary2018}). The important point 
is that the radio observations in the frequency range of 5 -- 10 GHz do not wholly resemble 
the soft X-ray or EUV, but they do largely imitate the photospheric magnetograms 
(\citealt{dabrowski2009}; \citealt{nita2004}; \citealt{white2011}). Since the magnetic 
complexity of an active region crucially determines the occurrence of intense flares and 
energetic CMEs (\citealt{priest2002}; \citealt{yashiro2005}), this white-paper emphasises
the importance of regular radio mapping of the Sun to reveal the magnetic characteristics 
of an eruptive region in line with the magnetogram data. 

\section{\fontsize{14}{14}\selectfont Solar Radio Mapping - Tracking the Evolution of 
Active \\ Regions}

Solar mapping observations with the Arecibo 12-m Radio Telescope were initiated in 
mid-December 2021. (The Arecibo Observatory is operated by the University of Central 
Florida under a cooperative agreement with the National Science Foundation, AST-1822073, 
and in alliance with Yang Enterprises and Universidad Ana G. M\'endez.) The 12-m radio 
telescope presently operates in the frequency ranges of 2.21 -- 2.34 GHz (S-band) and 
8.1 -- 9.2 GHz (X-band) and takes advantage of RFI protection from the Puerto Rico  
Coordination Zone (PRCZ at frequencies below 15 GHz), which covers Puerto Rico and 
nearby Puerto Rican islands. The 12-m antenna currently operates with room-temperature 
receiver systems and records dual polarization signals
({\it https://www.naic.edu/ao/scientist-user-portal/astronomy/12m-radio-telescope}).  
`East-west' raster scans of the Sun are taken routinely and maps are made from these. 
Figure 1 shows an example of a full image of the Sun at 8.6 GHz made on 25 February 
2022, along with the EUV image of the Sun observed by the {\it Atmospheric Imaging 
Assembly} (AIA) on board the {\it Solar Dynamics Observatory} (SDO) in the wavelength 
band of 193~\AA~(\citealt{pesnell2012}; \citealt{lemen2012SoPh}).
The spatial resolution of the 12-m telescope in the frequency band of 8.1 -- 9.2 GHz 
is limited to $\sim$10 arcmin. Nevertheless such maps provide a clear view of the 
emission brightness temperatures of active and quiet regions on the Sun, and regular 
monitoring is useful to follow the evolution of the emission of active regions. 
Since the 12-m radio telescope covers the frequency band of 8.1 -- 9.2 GHz, several 
maps are simultaneously made at frequency intervals of $\sim$170 MHz. The 
inter-comparison between these frequency bands is extremely useful and gives a handle 
on the mitigation and elimination of radio interference, should this be present.

\begin{figure}[!t]
\begin{center}
\includegraphics[height=14.6cm]{figure_2.ps}
\end{center}
\textbf{Figure 2:} Peak brightness (in kiloKelvin) of the Sun at 8.6~GHz obtained
from the Arecibo 12-m radio telescope plotted for dates from 13 December 2021 to 31 July 
2022, in the ascending phase of the current solar cycle 25 (top). In the middle and bottom 
panels 
hourly average EUV (EVE/SDO) and X-ray (GOES) fluxes are plotted for comparison. The three 
peaks marked on the radio plot with the letters `$a$', `$b$', and `$c$' are dominant, and 
correspond to strong emission from magnetically-active regions of `$\beta$-$\gamma$-$\delta$'
configuration.
\end{figure}

\begin{figure}[!b]
\begin{center}
\vspace{-0.2in}
\includegraphics[height=17.9cm]{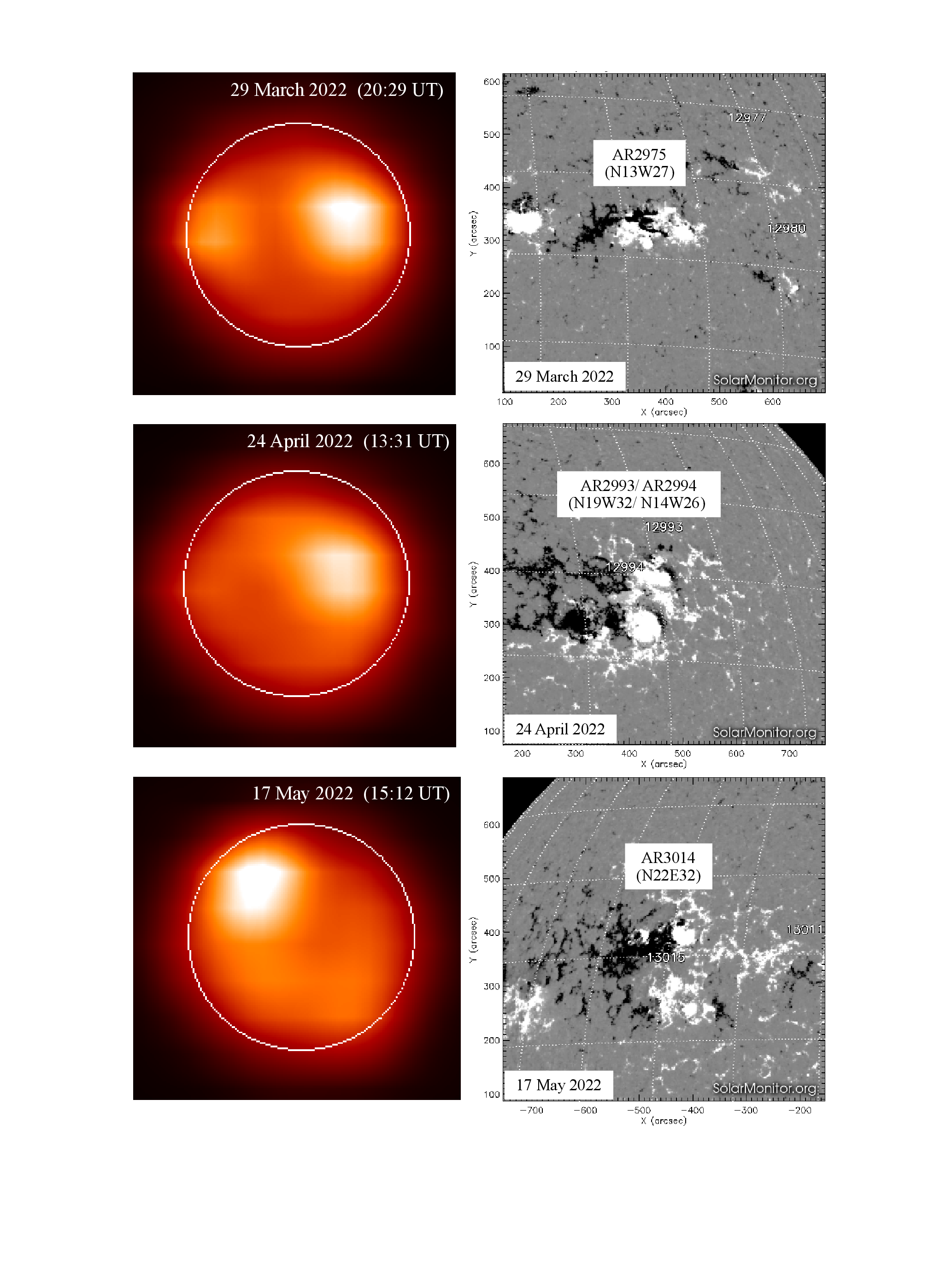}
\end{center}
\vspace{-0.7in}
\textbf{Figure 3:} Arecibo radio images at 8.6 GHz made on 29 March, 24 April, and 
17 May 2022, of peak brightness temperatures in the range of 13,000 -- 16,000~K and 
of magnetic configurations between $\beta$-$\gamma$ and $\beta$-$\gamma$-$\delta$ 
(left). On the right of each image is the corresponding HMI/SDO magnetogram of the 
bright emitting portion (courtesy of {\it SolarMonitor.org}).
\end{figure}

In Figure 2, the peak brightness of the Sun (in kiloKelvin, averaged over the 
10 arcmin beam) obtained from Arecibo's 8.6-GHz images is displayed for the 
period between 13 December 2021 and 31 July 2022, in the ascending phase of 
the current solar cycle 25, along with hourly-averaged EUV (0.1 -- 7 nm) 
irradiance of the Sun observed by the {\it Ultraviolet Variability Experiment} 
(EVE) on board SDO (\citealt{wood2012SoPh}) and X-ray (0.1 -- 0.8 nm) flux by 
the {\it Geostationary Operational Environmental Satellite} (GOES)-16 
({\it http://www.swpc.noaa.gov/Data/goes.html}). 
At Arecibo, depending on the other astrophysical observing programs, in a day 
minimum 2 or more images of the Sun are made. In the radio brightness 
plot, several systematic peaks are seen and each point on the plot
indicates the typical average peak brightness on the disk of the Sun. However, 
when the observation time coincided with a flare, it included the brightening 
corresponding to the particular phase of the flare. For example, one of the maps on 
day number 89 (30 March 2022) included the rising phase of the X-1.3 class flare, the 
first intense flare of the current solar cycle, and recorded a brightness temperature 
of $\sim$42,000~K (refer to the top panel in Figure 2). 

The strong radio emission from the Sun between day numbers 90 and 140, three peaks 
(indicated by letters `$a$', `$b$', and `$c$') show systematic increase and 
decrease and are much more prominent than the EUV and X-ray fluxes. These correspond 
to emission from multiple-pole magnetically-active, `$\beta$-$\gamma$-$\delta$', 
regions developed on the Sun, from where intense flares and energetic CME eruptions 
were observed (e.g., \citealt{jaeggli2016}; \citealt{yashiro2005}), and respectively 
correspond to active regions AR2975, AR2993/2994, and AR3014. In fact, the $b$-peak$'$s
AR2993/2994 were the return of AR2975, which appeared at the east limb of the Sun on 
15 April 2022, and in the subsequent rotation decayed to a less active state.  
For each peak, when the 
magnetic configuration of the active region attained $\beta$-$\gamma$ configuration, 
the brightness temperature increased to a level of $\sim$13,000~K. As the peak was 
approached, it reached $\sim$19,000 -- 21,000~K and the magnetic configuration 
developed to $\beta$-$\gamma$-$\delta$. The notable point is that all the 
M-class flares were produced when the brightness temperature was $\geq$13,000~K, 
whereas X-class flares occurred close to the peak at $\sim$20,000~K. Figure 3 shows 
the Arecibo radio images made at 8.6 GHz on 29 March, 24 April, and 17 May 2022, 
after the development of $\beta$-$\gamma$ magnetic configuration, of peak 
brightness temperatures in the range of 13,000 -- 16,000~K.  Alongside each
radio image,
the same day$'$s complex magnetogram of the bright emitting region is shown 
from the {\it Helioseismic Magnetic Imager} (HMI) (\citealt{scherrer2012})
on board the SDO space mission.

Figure 2 additionally reveals the interesting result that the brightness 
temperature of the quiet Sun, or low-activity region, is $\sim$8000~K as 
indicated by the dotted horizontal line in the top panel. This is likely 
in the middle of the chromosphere below 5000 km. With the 12-m radio 
telescope it would be interesting to detect ultra-intense flares and CMEs, 
rarely observed, when an active region attains the $\beta$-$\delta$ 
configuration.
However, the occurrence of this type of complex active region is infrequent 
and represented by $\sim$1\% of the sunspot population 
(e.g., \citealt{jaeggli2016}).  The regular solar imaging at the Arecibo 
will be essentially useful to track the formation of active regions, as 
well as the strong eruptions leading to extreme space-weather storms.

\subsection{\fontsize{12}{12}\selectfont Callisto: The Low-frequency Solar Radio 
Spectrometer}

The monitoring of the activity of the Sun at microwave frequencies is supported 
by the operation of the Callisto solar spectrometer at the Arecibo Observatory, 
in the low-frequency range of 15 -- 100 MHz. This allows a study of the radio 
emission associated with  
particle acceleration on the Sun and its related network of magnetic fields
({\it https://www.naic.edu/ao/ blog/ao-callisto-solar-radio-spectrometer}).
For example, the fast-drifting Type~III radio bursts are due to the acceleration 
of electrons along 
the open magnetic field lines in the corona into the solar wind and indicate 
the path of the energetic particles directed towards the Earth. 
The shock waves associated with CMEs are recorded as the slow-drifting Type~II 
bursts, which are extremely useful for understanding the initial propagation 
kinematics of CMEs and the shock-accelerated high-energy particles of space 
weather importance in the Sun to 1~AU space 
(\citealt{mano2000}, \citealt{gopal2013}). Apart from the above, other types of 
bursts, e.g., Type~I storms, Type~IV bursts (moving as well as stationary), U~type 
bursts, etc., potentially possess information of the different states of the 
magnetic configurations, as well as the heating of the plasma (\citealt{gopal2016}). 

\subsection{\fontsize{12}{12}\selectfont The Upgrade of 12-m Radio Telescope}

The 12-m radio telescope is currently being upgraded with a wideband, 2.3 -- 14 GHz, 
cooled front-end system, which will considerably enhance its sensitivity, as well as 
its frequency coverage. 
Since the new receiver also allows measurement of full-Stokes parameters, its extended 
bandwidth will provide highly accurate temporal measurements of polarization and dynamic 
spectra of the solar emission. This will be valuable for studying the evolution of the
magnetic-field configurations and plasma conditions in the current sheets of solar 
eruptions, as required for understanding the origin of space-weather events. Moreover, 
the mapping of the Sun over a wide frequency band will also provide the temporal and 
spatial evolution of the eruptive active regions at different layers between the
photosphere and the low corona.

In addition, the upgraded 12-m system will allow 
interplanetary scintillation (IPS) observations of compact background radio sources 
that can probe the ambient solar wind and structures within CMEs in the three-dimensional 
inner heliosphere, for regions inaccessible to spacecraft (e.g., \citealt{mano2010}). 
The above set of observations of high- and low-radio frequencies, plus IPS measurements, 
will provide a detailed view of the space weather events in the Sun-Earth space.

\subsection{\fontsize{12}{12}\selectfont The Involvement of Students in Space-Weather 
Studies}

Over the years, Arecibo Observatory has had a successful track record of 
implementing research and training programs to a large number of
students in the fields of astronomy, planetary science, and aeronomy.
A strong student program is required to engage undergraduate/graduate students, 
especially underrepresented groups, and provide them with an understanding of 
space-weather system science, including the origins of space weather on the Sun 
and Sun-Earth connections.  In the recent batch of NSF funded {\it Research 
Experience for Undergraduate} (REU) students at the Arecibo Observatory, a space 
weather research project was introduced.
The aim of the current NSF's Partnerships in Astronomy \& Astrophysics Research 
and Education (PAARE) program, `{\it Enhancing and Nurturing Careers in Astronomy 
with New Training Opportunities'} (ENCANTO), is to raise the number of successful 
underrepresented students from Puerto Rico and Florida in astronomy and planetary 
physics. ENCANTO represents an effort between the Arecibo Observatory and multiple 
higher educational institutions, the University of Central Florida and universities 
in Puerto Rico,
and it includes solar and space weather study projects 
({\it http://www.naic.edu/ao/encanto}).  Further, by involving more students, 
the Arecibo Observatory will become a thriving center of space weather studies 
that can increase the U.S. STEM workforce in a broader spectrum of space-weather 
education.

\section{\fontsize{14}{14}\selectfont Looking to the Future}

Space weather is increasingly becoming a key component in the day-to-day 
operation of several technological systems. 
Regular solar imaging at high frequencies has the potential for studying the
development of solar active regions and tracking the origin of extreme 
space-weather events.  It is shown that the centimetric-wavelength 
brightness temperature of an active region is the indicator of acceleration 
of electrons, i.e., magnetic field lines in the region above the photosphere, 
as well as close to the solar atmosphere. Particularly, in the radio X-band, 
a brightness temperature above a value of $\sim$13,000~K 
serves to identify an active region of strong eruptions. The great 
advantage of these measurements is that they can identify an eruptive region 
when it rotates close to the central meridian of the Sun, about one half to a 
day in advance, and predict the strongest flares and CMEs. Moreover, for a 
better forecasting strategy, such radio observations with a larger radio
telescope having the high-spatial 
and temporal resolution, along with a polarization capability, would have 
tremendous potential not only for following the magnetic activity of the 
eruptive site, but also for revealing the particle acceleration mechanism 
and additional exciting science. Designs are currently being developed for
a similar-sized replacement of the recently lost 305-m Arecibo radio 
telescope.

\vskip 20pt
\section*{\fontsize{14}{14}\selectfont Acknowledgments}

The Arecibo Observatory is operated by the University of Central Florida
under a cooperative agreement with the National Science Foundation
(AST-1822073), and in alliance with Universidad Ana G. M{\'e}ndez and Yang 
Enterprises, Inc. We acknowledge the EUV data from the UVE and images from 
the AIA and HMI on board the Solar Dynamics Observatory. The X-ray data sets 
have been obtained from the Geostationary Operational Environmental Satellite 
(GOES-16).

\vskip 10pt

\bibliography{astro_ph_version}{}
\bibliographystyle{plainnat}

\end{document}